\input harvmac
%\draftmode
%d  \baselineskip=14pt

\def \const {{\rm const}}

\def\a{\alpha}
\def\b{\beta}
\def\g{\gamma}

\def\r{\rho}

\def\te{\theta} \def \t {\theta}

\def\p{\phi}

\def \r {\rho}\def \g {\gamma} 
\def \ov {\over }\def \b {\beta}
\def \P {\Phi}  \def \const {{\rm const}}

\def\tchi{{\tilde{\chi}}}

\def\a{\alpha}
\def\b{\beta}
\def\N{{\cal{N}}}
\def\g{{\gamma}}

%%%%%%%%%%%%%%%%%%%%%%%%%%%%%%%%%%%

\def \t { t }

\lref\bhm{
 P.~Berglund, T.~Hubsch and D. Minic,
``de Sitter Spacetimes from Warped Compactifications of IIB String Theory,''
hep-th/0112079.
}

\lref\ghkt{
S.~S.~Gubser, C.~P.~Herzog, I.~R.~Klebanov and A.~A.~Tseytlin,
``Restoration of chiral symmetry: A supergravity perspective,''
JHEP {\bf 0105}, 028 (2001),
hep-th/0102172.
%%CITATION = HEP-TH 0102172;%%
}

\lref\KT{
I.~R.~Klebanov and A.~A.~Tseytlin,
``Gravity Duals of Supersymmetric $SU(N) \times
SU(N+M)$ Gauge Theories,''
{\it Nucl. Phys.} {\bf B578} (2000) 123,
hep-th/0002159.
%%CITATION = HEP-TH 0002159;%%
}

\lref\KS{I.~R.~Klebanov and M.~J.~Strassler,
``Supergravity and a Confining Gauge Theory:
Duality Cascades and $\chi$SB-Resolution of Naked
Singularities,''
{\it JHEP} {\bf 0008} (2000) 052,
hep-th/0007191.
%%CITATION = HEP-TH 0007191;%%
}

\lref\kuts{D.~Kutasov and D.~A.~Sahakyan,
"Comments on the Thermodynamics of Little String
Theory," 
{\it JHEP} {\bf 0102}, (2001) 021, hep-th/0012258.
}

\lref\ass{ O.~Aharony, E.~Schreiber and J.~Sonnenschein,
``Stable Non-Supersymmetric Supergravity Solutions from Deformations of the Maldacena-Nunez Background,''
hep-th/0201224.
}

\lref\Witt{E.~Witten,
``Anti-de Sitter space, thermal phase transition, and
confinement in gauge
theories,''
{\it Adv.~Theor.~Math.~Phys.} {\bf 2} (1998) 505,
hep-th/9803131.
%%CITATION = HEP-TH 9803131;%%
}

\lref\bt{A.~Buchel and  A.~A.~Tseytlin,
``Curved space resolution of singularity of fractional D3-branes on conifold,''
hep-th/0111017.
}

\lref\ms{J.~M.~Maldacena and A.~Strominger, 
"Semiclassical decay of near extremal fivebranes,"
{\it JHEP} {\bf 9712}, (1997) 008, hep-th/9710014.
}

\Title{\vbox
{\baselineskip 10pt
\hbox{   }\hbox{NSF-ITP-02-16}  }}
{\vbox{\vskip -30 true pt\centerline {
Gauge/gravity correspondence }
\medskip
\centerline {in accelerating universe
 }
\medskip
\vskip4pt }}
\vskip -20 true pt
\centerline{Alex Buchel }
\smallskip\smallskip
\centerline{ \it Institute for Theoretical
Physics,
University of California, Santa Barbara, CA
93106-4030}

\bigskip\bigskip
\centerline {\bf Abstract}
\baselineskip12pt
\noindent
\medskip
We discuss time-dependent backgrounds of type IIB
supergravity realizing gravitation duals of gauge theories
formulated in  de Sitter space-time as a tool
of embedding  de Sitter in a supergravity.
We show that only the gravitational duals to non-conformal
gauge theories are sensitive to a specific value of a
Hubble parameter. We consider two  nontrivial solutions
of this type: a gravity dual to six-dimensional
$(1,1)$ little string theory,
and to a four-dimensional cascading $SU(N+M)\times SU(N)$ supersymmetric
gauge theory (related to fractional D3-branes on a singular
conifold according to Klebanov et al),  in accelerating universe.
In both cases we argue that the IR singularity of the geometry
is regulated by the expansion  of the gauge theory
background space-time.

\bigskip

%%%%%%%%%%%%%%%%%%%%%%%%%%%%%%%%%%%%%%%%%%%%%%%%%%%%%%%%%
\Date{03/02}

%%%%%%%%%%%%%%%%%%%%%%%%%%%%%%%%%%%%%%%%%%%%%%%%%%%%%%%%%%%%%%%%%%%
\noblackbox
\baselineskip 16pt plus 2pt minus 2pt

\lref\fiv{
A.~Buchel, C.~P.~Herzog, I.~R.~Klebanov, L.~Pando Zayas and
A.~A.~Tseytlin,
``Non-extremal gravity duals for fractional D3-branes on the
conifold,''
{\it JHEP} {\bf 0104}, 033 (2001),
hep-th/0102105.
%%CITATION = HEP-TH 0102105;%%
}

\lref\agmoo{O.~Aharony, S.~S.~Gubser, J.~Maldacena, H.~Ooguri and Y.~Oz,
``Large N field theories, string theory and gravity,''
{\it Phys.\ Rept.\ }  {\bf 323}, 183 (2000),
hep-th/9905111.
%%CITATION = HEP-TH 9905111;%%
I.~R.~Klebanov,
``Introduction to the AdS/CFT correspondence,''
hep-th/0009139.
%%CITATION = HEP-TH 0009139;%%
}

\lref\kw{I.~R.~Klebanov and E.~Witten,
``Superconformal field theory on threebranes at a Calabi-Yau
singularity,''
{\it Nucl.\ Phys. B}\  {\bf 536}, 199 (1998)
hep-th/9807080.
%%CITATION = HEP-TH 9807080;%%
}

\lref\kn{I.~R.~Klebanov and N.~A.~Nekrasov,
``Gravity duals of fractional branes and logarithmic RG flow,''
{\it Nucl.\ Phys.\ B}  {\bf 574}, 263 (2000),
hep-th/9911096.
%%CITATION = HEP-TH 9911096;%%
}
\lref\ps{J.~Polchinski and M.~J.~Strassler,
``The string dual of a confining four-dimensional gauge theory,''
hep-th/0003136.
%%CITATION = HEP-TH 0003136;%%
}

\lref\btun{A.~Buchel and A.~A.~Tseytlin, unpublished notes.}

\lref\malda{
J.~M.~Maldacena,''The Large N Limit of 
Superconformal Field Theories and Supergravity,''
{\it Adv.~Theor.~Math.~Phys.} {\bf  2} 231 (1998),
hep-th/9711200. 
}

\lref\pgk{
S.~B.~Giddings, S.~Kachru and J.~Polchinski,
``Hierarchies from Fluxes in String Compactifications,''
hep-th/0105097.
}

\def \t {\tau}

%%%%%%%%%%%%%%%%%%%%%%%%%%%%%%%%%
\newsec{Introduction }
%%%%%%%%%%%%%%%%%%%%%%%%%%%%%%%%%%%
Gauge theory -- gravity duality\foot{For  reviews
and references see, e.g.,  \agmoo.} relates a gauge theory on the world
volume of
large number of D-branes to purely supergravity
backgrounds where the
branes are replaced by the corresponding fluxes.
In the simplest case the duality is realized \malda\
by a system of  $N$  D3-branes
in a flat type IIB string theory background. At small 't Hooft coupling
$g_s N\ll 1$, the system is best described by open strings and
realizes $SU(N)$ $\N=4$ supersymmetric
gauge theory. In the limit of strong 't  Hooft coupling this gauge theory
has a perturbative description as 
type IIB supergravity compactified on
$AdS_5\times S^5$,
with $N$ units of the RR 5-form flux through the $S^5$.
If this is 
a genuine equivalence, then  phenomena observed on the
gauge theory side should have  dual description  in string theory
on  $AdS_5\times S^5$.
In particular,  {\it any} deformation on
the gauge theory visible in the
large $N$ limit should have a counterpart
in the dual gravitational
description, and vice versa.

As in \bt\  we use ``deformations'' in a generalized sense.
For example, Klebanov-Witten duality \kw\
describing  regular D3-branes placed at  a conical singularity
in type IIB string theory can be thought of as
a $Z_2$ orbifold of the original duality of Maldacena \malda\
along with a certain mass deformation that leaves only a quarter of
the original supersymmetries unbroken.
One could go a step further and consider deformations of a
background space-time in which one formulates gauge dynamics.
In \bt\ a  gauge-gravity correspondence was considered
in which  Minkowski background space-time of the
Klebanov-Strassler (KS) \KS\ cascading gauge theory
was replaced with $R\times S^3$ or (in a Euclidean case) $S^4$.
It was  argued there that the curvature of the background geometry
provided an infrared cutoff on the gauge theory dynamics
and  resolved the Klebanov-Tseytlin (KT) \KT\ naked singularity.

A natural extension to proposal of \bt\ is to ask the question: what the
gauge-gravity duality of Maldacena would look like
when the gauge theory space-time background is de Sitter? 
This is a perfectly valid ``deformation'' of the 
gauge theory background  where one ``turns on'' 
a Hubble parameter. And thus, provided the original 
gauge-gravity correspondence was exact, one should be able to map this 
deformation onto the dual supergravity. 
In this paper we describe such a map.
We would like to emphasize that much like in the original 
Maldacena correspondence, the gauge theory space-time is not dynamical
on the gauge theory side of the correspondence. In other words,
on the gauge theory side of the correspondence we completely 
neglect the backreaction of the gauge theory dynamics on the background, we  
ignore background fluctuations as well.
The story on the  gravity side of the correspondence is drastically different:
here, as in the original Maldacena correspondence, what was the gauge theory 
background becomes a part of a dynamical type IIB supergravity background.  
Thus, finding a gravity dual to a gauge theory on a (decoupled) dS background 
would provide an embedding of this space-time into dynamical supergravity.
Put differently: we want to view ``cosmological'' deformation
of the gauge-gravity correspondence as a  
tool of embedding a de Sitter space-time into a supergravity\foot{
Related ideas of realizing de Sitter gravity in warped compactifications 
of type IIB string theory were discussed in \bhm\ . }.

The paper is organized as follows.
In the next section we describe a  motivation
for a time-dependent metric ansatz of type IIB supergravity background
dual to a gauge theory in accelerating universe.
We observe that  de Sitter deformation applied to the 
$\N=4$ $SU(N)$ supersymmetric Yang-Mills gauge/gravity correspondence
does not give rise to a different geometry on the dual supergravity 
side: all we get is a de Sitter slicing of the AdS factor in the 
original Maldacena duality. Nonetheless, we expect 
physically the deformed gauge 
theory to be rather different from the undeformed 
one. 
In particular, because of the conformal coupling of the gauge theory
scalars to the scalar curvature, in the $H\ne 0$ ($H$
 is the expansion rate of the universe) case the SYM 
would not have a moduli space\foot{A similar phenomenon for the 
$R\times S^3$ deformation of the $\N=4$ SYM was emphasized in \bt\ .
}$^,$\foot{It would be very interesting to study de Sitter deformations of 
gauge theories from purely field theoretical perspective. 
In this paper we focus on the supergravity part 
of the de Sitter deformed gauge-gravity correspondence.}.    
We further show that conformal gauge theories are the only
examples for which  supergravity duals for nonzero Hubble parameter
$H$ are related by some coordinate reparametrization to their $H=0$
supergravity duals.

We then move on to consider non-conformal examples in
sections 3 and 4. In section 3 we present supergravity dual to
$(1,1)$ little string theory (LST) in an inflationary 
patch of the $dS_6$. The $H=0$ solution reproduces the BPS 
system of $N\gg 1$ NS5-branes,
and thus has curvature singularity (in the Einstein frame) 
at the branes core. From the dual
gauge theory perspective this singularity is generated by the
zero modes of the $d=6$ SYM theory, which is the infrared limit
of $(1,1)$ LST. We explicitly demonstrate that de Sitter deformation 
of LST  regulates this curvature singularity. 

In section 4 we briefly discuss gravitational
dual to  Klebanov-Tseytlin/Klebanov-Strassler (KT/KS) 
cascading gauge theory \refs{\KT,\KS} in accelerating universe. 
We show that the  KT deformation 
is related by a Wick rotation plus some scaling of the
KT gauge theory on $S^4$ previously considered in \bt\ . 
Thus the infrared singularity of the extremal KT geometry 
is resolved for $H\ne 0$ as explained in \bt\ .
We conclude in section 5.

%%%%%%%%%%%%%%%%%%%%%%%%%%%%%%%%%%%%%%%%%%%%%%%%%%%%%%%
\newsec{Supergravity duals of gauge theories in accelerating universe}
%%%%%%%%%%%%%%%%%%%%%%%%%%%%%%%%%%%%%%%%%%%%%%%%%%
We mentioned in the introduction that  given the original
gauge-gravity duality of Maldacena there is  a  simple way to embed
dS  space-time into supergravity. The reason for this is
that  since we can deform a background space-time of the
gauge theory from Minkowski to  a flat Robertson-Walker universe 
by simply ``turning on'' a Hubble parameter,
we should be able to do this in the supergravity dual to
this gauge theory.

Typically, in a  gauge-gravity correspondence
the dual supergravity metric\foot{We always work in
the Einstein frame.} can be written  as
\eqn\ormett{
ds_{10E}=c_1^2 (dM_d)^2+ c_2^2 dr^2 +(d\mu_{9-d})^2\,,
}
where $M_d$ is  a  $d$-dimensional Minkowski space-time,
which is related to the space-time background of the
dual gauge theory, and  $\mu_{9-d}$ (for a fixed $r$)
is a compact $(9-d)$-dimensional Riemannian manifold
that encodes the gauge theory dynamics at energy scale
$E\sim \rho$ with $c_2dr\sim {d\rho\over \rho}$ as
$\rho\to \infty$. From now on we consider only the cases
where $c_i$ depend only on $r$\foot{This is not always the case,
as for example in  Polchinski-Strassler gauge-gravity
correspondence \ps\ .}. The metric on $M_d$
does not depend on the angles of $\mu_{9-d}$
while $(d\mu_{9-d})^2$ does not depend on the $M_d$
coordinates, though both $M_d$ and $\mu_d$ can  have
explicit $r$ dependence\foot{The
examples where $(dM_d)^2$ has an $r$ dependence correspond to
gauge theories formulated on compact manifolds as in \refs{\Witt,\bt}. }.
It seems natural to assume that such ``separation of variables''
would hold even when we start deforming the
gauge theory space-time $M_d$. Specifically, taking the
$d$-dimensional gauge theory in an accelerating universe
\eqn\rw{
(ds_d^H)^2=-dt^2+e^{2 H t }d\bar{x}^2\,,
}
which for $H=0$ has a dual supergravity background with
the metric \ormett\ ,  we assume the metric ansatz of the
dual supergravity  for general $H$ to be
\eqn\genmd{
(dM_d)^2\to (dM_d^H)^2\equiv  (ds_d^H)^2\,.
}
In the orthonormal frame 
\eqn\frame{
e^1=c_1 dt\,,\qquad e^{i+1}=e^{H t} c_1 dx_i\,,\ i=1\cdots d-1\,,\qquad
e^{d+1}=c_2 dr\,,
}
\eqn\frametwo{
e^j\,,\ j=d+2,\cdots,10\quad {\rm such\ that\quad }    e^j e^j\equiv (d\mu_{9-d})^2\,,
}
the ricci components of the metric
are  time independent; so 
in this frame the supergravity fluxes and the dilaton 
would be time-independent
as well.

We begin explicit examples by considering the case of the 
$H\ne 0$ deformation of the gauge-gravity correspondence discussed in
\kw\ where the gauge theory is conformal, namely D3 branes
at a conical singularity\foot{There is an obvious generalization
to $AdS_5\times S^5$, and other conformal cases.}. We observe 
that the dual supergravity background for $H\ne 0$ still remains  
$AdS_5\times S^5$: the only difference is that now we are doing 
a de Sitter slicing of the AdS factor in the metric. 

Type IIB equations of motion can be solved analytically in this case.
We find
\eqn\dthree{
ds_{10}^2=\r^2\left(-dt^2+e^{2H t}d\bar{x}^2\right)
+{ L^2 d\r^2\over  L^2 H^2 +\r^2}+ L^2 ds_{T^{1,1}}^2\,,
}
where $(ds_{T^{1,1}})^2$ is the standard metric on 
$T^{1,1}=(SU(2)\times SU(2))/U(1)$ and
\eqn\defl{
L^4=4\pi g_s N(\a') {27\over 16}\,,
}
with $N$ being the number of  D3 branes.
The metric \dthree\ is supported by the following 5-form
flux
\eqn\flux{
F_5={\cal F}_5+\star{\cal F}_5\,,\qquad {\cal F}_5={- L^4} 
d{\rm vol}_{T^{1,1}}\,.
}
Above solution is related
by a  coordinate transformation to the extremal ($H=0$) D3 brane solution.
Indeed, first introduce
\eqn\timec{
\tau={1\over H} e^{-H t}\,.
}
Then the change of variables that do the job is
\eqn\chao{
r={\rho\over H\tau}=\rho e^{H t}\,,
}
\eqn\chat{
d\tilde{t}=-{\sqrt{ L^2 H^2 +\r^2}\over \r}d\tau-{ L^2 H^2 \tau\over
\sqrt{ L^2 H^2 +\r^2}} {d\r\over \r^2}\,,
}
where $r$ and $\tilde{t}$ are the radial and the time coordinates
of the $H=0$ solution.
Note that in  \chat\ $d^2 \tilde{t}=0$, so this equation can indeed
be integrated
\eqn\chatint{
\tilde{t}={\sqrt{ L^2 H^2 +\r^2} e^{-H t}\over H\r}\,.
}
From  the coordinate transformations \chao\ , \chatint\ 
we see that de Sitter slicing of $AdS_5$, as in \dthree\ ,  
covers ``half'' ($\tilde{t}\ge 0$ region) of its  Poincare patch. 
It is easy to see that this slicing can be obtained from the 
analytical continuation (along with some scaling limits) of the 
Euclidean $AdS$ in the ``hyperboloid'' 
parametrization\foot{This is a Euclidean 
$AdS_{d+1}$ parametrization where the constant radial slice
is $S^d$.} . Really, 
for the $AdS_{d+1}$ the metric in this parametrization is given by
\eqn\changemet{
ds^2_{AdS_{d+1}}=\sinh^2\r \left(dS^d\right)^2+d\r^2
=\sinh^2\r \left(d\t^2+\sin^2\t \left(dS^{d-1}\right)\right)^2+d\r^2\,.
} 
Now Wick rotation of \changemet\ $\t\to i\t$ and the 
``decompactification'' limit of $S^{d-1}$, $(dS^{d-1})^2\to (dR^{d-1})^2$,
along with $\t\gg 1$ give
\eqn\chatwo{  
ds^2_{AdS_{d+1}}
=\sinh^2\r \left(-d\t^2+e^{2\t} \left(dR^{d-1}\right)\right)^2+d\r^2
}
$$
=r^2 \left(-d\t^2+e^{2\t} \left(dR^{d-1}\right)\right)^2+{dr^2\over 1+r^2}\,, 
$$
where $r\equiv \sinh\r$.
Thus,  coordinate transformations  \chao\ , \chatint\ 
must be  
represented by the (corresponding scaling limit of the) Wick rotation of
local coordinate transformations relating Poincare and ``hyperboloid''
parametrizations of the Euclidean $AdS$.

In the rest of this section we address the question  when does the $H\ne 0$
deformation of a given gauge-gravity duality
is related to the original ($H=0$) correspondence
by some change of variables, as in the case above.
We will argue that this is so only when the gauge theory in the
duality correspondence is conformal.
Let
\eqn\mettzero{
(ds_{10E}^0)^2=(c_1^0)^2 \left(-dt^2+d\bar{x}^2\right)+ (c_2^0)^2 dr^2
+(d\mu_{9-d}^0)^2
}
be a supergravity metric in the original\foot{That is a 
gauge theory is formulated in  Minkowski space-time.} gauge-gravity 
correspondence, and
\eqn\metth{
(ds_{10E}^H)^2=(c_1)^2 \left(-d\tau^2
+e^{2 H \tau} d\bar{x}^2\right)+ (c_2)^2 d\r^2
+(d\mu_{9-d})^2
}
is the metric corresponding to its $H\ne 0$ deformation.
We want to know when \metth\  is related by some coordinate
reparametrization  to \mettzero\ .
Replacing $\tau\to {1\over H}e^{-H \tau}$ in \metth\  we get
\eqn\mettht{
(ds_{10E}^H)^2={c_1^2\over H^2\tau^2}\left(-d\tau^2 +d\bar{x}^2\right)
+c_2^2 d\r^2 +(d\mu_{9-d})^2\,.
}
Let's ignore for now  the internal piece of the metric.
Comparing $d\bar{x}^2$ pieces of the metric
in \mettht\ and \mettzero\ we see that
\eqn\cone{
c_1^0(r)={c_1(\r)\over H\tau}\,,
}
so that
\eqn\dr{
dr={[c_1(\r)]'\tau d\r-c_1(\r)d\tau\over H\tau^2 [c_1^0(r)]'}\,.
}
Taking the most general ansatz for $dt$
\eqn\dt{
dt=g_1(\r,\t) d\tau+g_2(\r,\t) d\r\,,
}
and matching \mettht\ and \mettzero\ we find
\eqn\gone{
g_1(\r,\t)={c_1(\r)\sqrt{[c_2^0(r)]^2+\tau^2 [c_1^0(r)']^2}\over
c_1^0(r)H\tau^2 [c_1^0(r)]' }\,,
}
\eqn\gtwo{
g_2(\r,\t)=-{[c_2^0(r)]^2\ [c_1(\r)]'\over H\tau c_1^0(r) [c_1^0(r)]'
\sqrt{[c_2^0(r)]^2+\tau^2 [c_1^0(r)']^2}}\,,
}
plus we have a constraint\foot{We are  assuming that the $c_1^0$
warp factor in \mettzero\ is nontrivial, that is not a constant.}
\eqn\constr{
0=-[c_2^0(r)]^2
[c_1(\r)']^2+c_2(\r)^2 H^2 [c_2^0(r)]^2+c_2(\r)^2 H^2\tau^2
[c_1^0(r)']^2\,.
}
Since we should be able to integrate \dt\
\eqn\intc{
d^2t\equiv 0\,.
}
It turns out, given above expressions we can rewrite \intc\
as
\eqn\res{
0={d\over dr}\left[{[c_1^0(r)]'\over c_1^0(r) c_2^0(r)}\right]\,.
}
Without loss of generality we can assume that in the
original duality\foot{This fixes  an arbitrary choice
of a  radial coordinate in \mettzero\ . }
\eqn\coneo{
c_1^0(r)=r\,.
}
From \res\ we find then
\eqn\ctwoo{
c_2^0={L\over r}\,,
}
where $L$ is some constant.
Finally, the only way  $(d\mu_{d-9})^2$ and  $(d\mu_{d-9}^0)^2$
could ever match is when they are independent of $\r$ and $r$
correspondingly. Thus we conclude that the metric \mettzero\
is actually
\eqn\final{
(ds_{10E}^0)=(ds_{AdS_{d+1}})^2+(d{\mu}_{9-d})^2\,,
}
where the metric on $\mu_{9-d}$ does not depend on the $AdS_{d+1}$
radial coordinate.
The AdS factor in \final\ points to the conformal invariance
of the dual gauge theory.

Above discussion suggests that for the
embedding of a de Sitter space-time  in supergravity 
we should look for deformations of
gauge/gravity duality where the gauge theory is not conformal.
We will present explicit examples of such deformations
in the next two sections.

%%%%%%%%%%%%%%%%%%%%%%%%%%%%%%%%%%%%%%%%%%%%%%%%%%%%%%%
\newsec{$(1,1)$ LST in accelerating universe and the IR singularity
resolution by inflation}
%%%%%%%%%%%%%%%%%%%%%%%%%%%%%%%%%%%%%%%%%%%%%%%%%%
In this section we describe $H\ne 0$
deformations of the $(1,1)$ little string theory,
realized on the world-volume of NS5 branes in type IIB 
string theory.  
The effective infrared description of the LST is in terms 
of $d=6$
 $\N=2$ supersymmetric Yang-Mills theory. As this gauge theory 
is not conformal, we expect to get a nontrivial embedding of $dS_6$
from its $H\ne 0$ deformation.

In the extremal case, $H=0$, the supergravity 
approximation breaks down near the core of the branes. 
This curvature singularity can be thought of as being generated 
by the zero modes of the IR free $d=6$ SYM. Since a Hubble parameter provides
an infrared cutoff on the dynamics of the theory, we expect that it 
should  regulate the curvature singularity of the extremal background. 
We argue that this is indeed so. 

We take the following ansatz for the metric of LST holographic 
dual in the inflationary patch of the $dS_6$
\eqn\metlittle{
(ds_{10E})^2=c_1^2 \left(-dt^2+e^{2H t} d\bar{x}^2\right)
+c_2^2 d\r^2+ {c_3^2\over 4}\left(g_1^2+g_2^2+g_3^2\right)\,,
} 
where $c_i=c_i(\r)$, and $g_i$ are the $SU(2)$ left-invariant one-forms
\eqn\oneforms{
g_1=\cos\phi\ d\theta+\sin\phi\sin\theta\ d\psi\,, 
}
$$
g_2=\sin\phi\ d\theta-\cos\phi\sin\theta\ d\psi\,, 
$$
$$
g_3=d\phi+\cos\theta\ d\psi\,.
$$
We assume the dilaton $\Phi\equiv \ln g_s$ to be a function of $\r$ only 
and the same NS-NS 3-form fluxes as in the extremal case
\eqn\fluxl{
H_3=n\ g_1\wedge g_2\wedge g_3\,,
}  
where $n$ is related to the  number of NS5 branes.  
Solving type IIB supergravity equations we get
\eqn\dres{
0=\left[{g_s'c_1^6 c_3^3\over g_s c_2}\right]'+{32 n^2 c_1^6 c_2 
\over c_3^3 g_s}\,,
}
\eqn\oneres{
0=\left[c_1'c_1^5c_3^3\over c_2\right]'-{c_1^4c_2\over g_s c_3^3}
\left(5 H^2 g_s c_3^6 +8 n^2 c_1^2\right)\,,
}
\eqn\threeres{
0=\left[c_3'c_3^2c_1^6\over c_2\right]'-{2 c_1^6 c_2\over g_s c_3^3}
\left(g_s c_3^4 -12 n^2\right)\,,
}
along with the first order constraint
\eqn\constres{
0={12 g_s^2 c_3^4\over c_1^4}\left[c_3 c_1\right]'\left[c_1^5 c_3\right]' -c_3^6 c_1^2 
\left(g_s'\right)^2+4 g_s c_2^2 \left(16 n^2 c_1^2-3g_s c_3^4 \left[c_1^2+5 H^2 c_3^2\right]\right)\,.
}
It is consistent with \dres\ - \constres\  to choose an ansatz 
for the warp factors $c_i$  similar to the extremal NS5 brane 
solution 
\eqn\ansa{
c_1=f g_s^{-1/4}\,,\qquad c_2=c_3=2 n^{1/2} g_s^{-1/4}\,.
}
We will end up with the following equations for  $f,g_s$
\eqn\edil{
0=\left[g_s' f^6\over g_s^3\right]'+{2   f^6 \over
 g_s^2 }\,,
}
\eqn\cmixed{
0=\left[\left({g_s\over f^4}\right)' {f^{10}\over g_s^3}\right]'
+{2 f^4 (40 H^2 n +
f^2)
\over  g_s^2}\,,
}
along with a first order constraint 
\eqn\const{
0=g_s^2 \left(60 H^2 n +2 f^2\right)-\left(15 g_s^2 [f']^2-12 g_s' f' g_s f 
+2 [g_s']^2 f^2\right)\,.
}  
Though we can not solve  \edil\ - \const\ analytically, it is 
straightforward to exhibit a smooth  solution. 
Really, a smooth solution as $\r\to 0$ is 
\eqn\gszero{
g_s=g_0\left(1-{1\over 7}\r^2+{5\over 378}\r^4+O(\r^6) \right)\,,
}      
\eqn\fzero{
f=2 H n^{1/2}\left(\r-{1\over 63}\r^3+{1\over 1470}\r^5+O(\r^7)\right)\,,
}
where $g_0$ is an integration constant 
related to the  string coupling.   
As $\r\to \infty$ we rather find\foot{These asymptotics can also 
be verified by numerical integration.} 
\eqn\gsinfty{
g_s\to  g_0\r^{3/4} e^{-\r}\,,\qquad f\to H n^{1/2}\sqrt{20\r}\,.
}
Note that curvature of \metlittle\ can be maintained arbitrarily  small 
by taking $g_0$ small. Thus the $H\ne 0$ deformation indeed 
regulates the strong curvature region of the extremal 
NS5 brane background.        
On the other hand, from \gsinfty\ we see that turning on a Hubble parameter 
induces a logarithmic 
correction to  the asymptotically linear dilaton background of the 
extremal NS5 branes. This should be contrasted with the finite energy density  
regularization of this geometry where one still recovers 
asymptotically linear dilaton \ms\ . 

From the above analysis it appears that given the Hubble 
parameter $H$, and for a fixed number of NS5 branes, 
there is a one parameter family of the LST de Sitter deformations, 
characterized by $g_0$. Furthemore,  it is $g_0$ and not $H$ 
that controlls curvature of the geometry \metlittle\ .   
This is suprising, as 
LST does not have any continuous coupling constant. Also,
physically, we expect that supergravity approxiation describing 
deformed LST should break down for sufficiently small $H$ 
(in string units), as this theory should still be 
weakly coupled at low energies. This suggests that $g_0$ 
can not be a free parameter. In what follows we argue that this is 
indeed so. We find that 
\eqn\gH{
g_0\sim 1/H^4\,, 
}
so that small $g_0$ 
(necessary for the validity of the supergravity description)
corresponds to a large Hubble parameter in string units, and thus 
the full picture is consistent with the general lore for the absense 
of a dual supergravity description to a  weakly coupled gauge
theory. Before we proceed with an argument for \gH\ we would like 
to mention that a somewhat similar phenomenon occurs in the near 
extremal deformation of the NS5 branes \ms\ . Really, 
the near extremal deformation of LST is characterized by a single 
parameter\foot{Classically, the temperature of the 
LST is independent of the energy density.}, namely the  energy density $\mu$. 
On the other hand,  its holographic dual naively  has two parameters:
$r_0$ (the location of the black five-branes horizon) and 
$g_h$ (the value of the string coupling at the horizon). 
It turns out that by a simple change of a radial coordinate 
the background geometry of the near extremal NS5 branes
can be shown to depend only on a combination $r_0^2/g_h^2$,
which can be further identified with the energy density 
above the extremality in string units $\mu$ \ms\ .   
In our case, though we described a two-parameter family $\{g_0,H\}$ 
of the regular solutions of \edil\ -\const\ , 
$H$ dependence of the geometry can also be eliminated  
by redefining the time coordinate $t\to \t\equiv 1/H e^{-H t}$. 
This is not very illuminating, as in doing so we are changing the 
reference energy scale from the LST perspective.
Rather, we continue measuring all energies in string units.
To relate $g_0$ and $H$ we study propagation of a 
minimally coupled scalar in the background \metlittle\ and on the 
NS5 brane probe.
Specifically, consider a massless scalar  $\chi$ 
minimally coupled to the Einstein metric \metlittle\ 
with zero angular momentum on $S^3$. The corresponding wave equation 
is 
\eqn\waveG{
0=-\partial_t\left[e^{5 H t}\partial_t\chi\right]
+e^{3 H t}\partial_i^2\chi+{e^{5 H t} g_s^2(\r)\over 4 n f^4(\r)}\ 
\partial_\r\left[{f^6(\r)\over g_s^2(\r)}\partial_\r\chi\right]\,,
}
where $i$ denotes the spatial directions of the NS5 brane. 
The last term in 
\waveG\ can be interpreted as a $\r$-dependent  mass term operator 
on the LST space-time. Using  \edil\ -\const\ we can explicitly
factor out $\{g_0,H\}$ dependence of this operator
\eqn\wavwG{
{e^{5 H t} g_s^2(\r)\over 4 n f^4(\r)}\ 
\partial_\r\left[{f^6(\r)\over g_s^2(\r)}\partial_\r\cdots \right]
\equiv e^{5 H t} H^2 {\cal O}(\r)[\cdots ]\,.
}      
Assuming the factorized dependence of $\chi$ on $\rho$
\eqn\fact{
\chi(t,\bar{x};\r)=\kappa(\r) \tilde{\chi}(t,\bar{x})\,, 
}
we get from \waveG\ 
\eqn\waveGt{
0=-\partial_t\left[e^{5 H t}\partial_t\tchi\right]
+e^{3 H t}\partial_i^2\tchi+e^{5 H t} H^2 \lambda(\r) \tchi\,, 
}
where $\lambda(\r)\equiv 1/\kappa(\r) {\cal O(\r)}[\kappa(\r)]$.
As in the original gauge-gravity correspondence of Maldacena
we would like to interpret $\r$ as a (measured in string units) 
holographic RG scale. Thus the dynamics of $\tchi$ should be 
qualitatively similar to the dynamics of the generically {\it massive}  
scalar $\eta$ propagating along a probe NS5 brane oriented along $\{t,\bar{x}\}$, and sitting at a 
fixed radial coordinate $\r$. 
With a scalar $\eta$ minimally coupled to the induced Einstein frame 
metric on the probe we find it's wave equation to be  
\eqn\etasc{
0=-\partial_t\left[e^{5 H t}\partial_t\eta\right]
+e^{3 H t}\partial_i^2\eta+e^{5 H t} m^2 g_s^{-1/2}(\r) \eta\,, 
}
where $m$ is a constant mass of $\eta$. Extracting the $g_0$ 
dependence of the last term in \etasc\ and comparing with \waveGt\ 
we are led to the identification \gH\
$$
g_0^{-1/2}\sim H^2\,.
$$

We believe that above arguments relating $g_0$ and $H$ 
are  qualitatively correct, and apparently 
lead to the expected physical picture. It is important to find 
a more precise understanding of this relation, or in other words 
the map between the supergravity parameters $\{g_0,H\}$ and their 
LST dual. This will likely  require an understanding of how to 
measure energies in non-static, asymptotically non-flat 
supergravity backgrounds.  
Finally, it is well known that string propagation in the throat geometry 
of the near extremal NS5 branes  corresponds to an exact
conformal field theory\foot{See, for example, \kuts\ . }.   
It would be interesting to see whether there is a CFT description
of the (1,1) LST in the de Sitter background presented here.

%%%%%%%%%%%%%%%%%%%%%%%%%%%%%%%%%%%%%%%%%%%%%%%%%%%%%%%
\newsec{de Sitter deformations of the KT/KS backgrounds }
%%%%%%%%%%%%%%%%%%%%%%%%%%%%%%%%%%%%%%%%%%%%%%%%%%

Our aim in this section will be to explore
dS embedding in the supergravity  in the context of 
the corresponding deformation of the KT model \KT\ .
Here, the conformal invariance of the gauge theory on the
D3 branes at a conical singularity \kw\ is broken by adding
fractional D3 branes \kn\ . We also comment on the de Sitter 
deformation of the KS background \KS\ .

We will start with  the same ansatz  as in \refs{\KT}
 and simply
replace 1+3 ``longitudinal'' directions by
the Robertson-Walker metric with flat spacelike
hypersurfaces
\eqn\frw{
(dM_4^H)^2=-dt^2+e^{2 H t} d\bar{x}^2\,.
}
As we show\foot{I would like to thank 
Arkady Tseytlin for pointing this out.}, 
there will be a direct relation to the KT model
on $S^4$ considered in \refs{\bt}.

As in \KT\ we will  impose the requirement
that
the background has abelian  symmetry associated with
the $U(1)$
fiber of $T^{1,1}$  as we will consider a phase where chiral
symmetry is restored\foot{Much like in the case of the LST on the de Sitter 
space-time, we expect  the Hubble scale to realize an IR cutoff on the 
gauge theory dynamics. Thus for sufficiently high $H$ (which we take 
to be the case in this section) we expect 
restoration of the chiral symmetry in the dual gauge theory.}.
Our  general ansatz for a 10-d 
 Einstein-frame metric will involve
3 functions $y,z$ and $w$  of  radial coordinate $u$ \foot{
For $H=0$ this
 metric
can  be brought into a more familiar
form
$
ds^2_{10E} =  h^{-1/2}(r)  (dM_4^{H=0})^2
+  h^{1/2}(r)   [  dr^2
+ r^2  ds^2_5] \ , $
where
$
 h=  e^{-4z} ,\    r  = e^{y  + w
 } ,\  e^{10y}
du^2  =  dr^2.$
When $w=0$ and $e^{4y}=r^4= { 1 \ov 4u} $,
the transverse 6-d space is the standard conifold
with $M_5= T^{1,1}$.
Small  $u$  thus  corresponds to large distances in 5-d
 and vice versa. In the $AdS_5$ region
 large $u$ is near the origin of
$AdS_5$ space, while $u=0$ is its boundary.}
\eqn\mett{
ds^2_{10E} =  e^{2z} (dM_4^H)^2
+ e^{-2z}  [e^{10y} du^2 + e^{2y} (dM_5)^2] \,.    }
Here   $M_5$ is a deformation of the $T^{1,1}$ metric
\eqn\mmm{
(dM_5)^2 =  e^{ -8w}  e_{\psi}^2 +  e^{ 2w}
\big(e_{\theta_1}^2+e_{\phi_1}^2 +
e_{\theta_2}^2+e_{\phi_2}^2\big)  \ , }
$$
 e_{\psi} =  {1\ov 3} (d\psi +  \cos \theta_1 d\phi_1
 +  \cos \theta_2 d\phi_2)  \  ,
 \quad  e_{\theta_i}={1\ov \sqrt 6} d\theta_i\ ,  \quad
  e_{\phi_i}=
{1\ov \sqrt 6} \sin\theta_id\phi_i \ .
$$

As for the matter fields,
we  will assume that the dilaton $\P$ may depend on $u$, and
our ansatz for the $p$-form  fields
 will be   exactly
as in the extremal KT case \KT\ and in \refs{\fiv,\bt}:
\eqn\har{
F_3 = \   P
e_\psi \wedge
( e_{\theta_1} \wedge e_{\phi_1} -
e_{\theta_2} \wedge e_{\phi_2})\ ,
\ \ \ \ \ \ \
B_2  = \    f(u)
( e_{\theta_1} \wedge e_{\phi_1} -
e_{\theta_2} \wedge e_{\phi_2})
 \ ,
}
\eqn\fiff{
F_5= {\cal F}+*
{\cal F}\
, \quad  \ \ \ \ {\cal F} =
K(u)
e_{\psi}\wedge e_{\te_1} \wedge
e_{\p_1} \wedge
e_{\te_2}\wedge e_{\p_2}\  , \ \ \ \ \ \ \
 K (u)  = Q + 2 P f (u) \ , }
where, as  in \KT, the  expression for $K$
follows from the Bianchi identity for the
5-form. The constants $Q$ and $P$ are proportional
to the numbers of standard and fractional
D3-branes.

We could now directly derive the corresponding system
of type IIB supergravity equations of motion describing the
radial evolution of the five unknown functions of
$u$:
$y,z,w,K,\Phi$. A better approach is to notice that the background we
consider here could be obtained from the KT model on $S^4$ discussed in
\bt\ . Really, the only difference of our case with the $S^4$
compactification of \bt\ is the replacement of the ``longitudinal''
directions \frw\ with $(dS^4)^2$
\eqn\tris{
(dM_4^H)^2\to  (dM_4)^2\equiv (dS^4)^2 =
 d \a^2 + \sin^2\a \ [d \b^2 + \sin^2 \b\  ( d \g^2
+ \sin^2
\g\  d \delta^2   ) ]  \,.  }
Now, Wick rotation of \tris\ $\a\to i\a$ and the
scaling limit on $S^3$ parameterized
by $\b,\gamma,\delta$ $(dS^3)^2\to d\bar{x}^2$ along with $\a\gg 1$ gives
\eqn\etris{
(dS^4)^2\to -da^2+e^{2\a} d\bar{x}^2\,,
}
which is precisely \frw\ with $H=1$.
Thus, the resulting  equations\foot{We also obtained these 
equations directly in the background \mett\ .} are just the straightforward
modification of (4.7)-(4.12) of \bt\ 
\eqn\yy{ 10y'' - 8 e^{8y} (6 e^{-2w} - e^{-12 w})  -
 30  H^2  e^{10y-4z }
 + \P''
=0 \ ,
}
\eqn\uw{
10w'' - 12 e^{8y} ( e^{-2w} - e^{-12 w})   - \P''
=0 \ , }
\eqn\pp{
\P''    + e^{-\P + 4z - 4y-4w} ({K'^2\ov 4 P^2} -
 e^{2 \P + 8 y+8w} P^2)=0 \ ,
}
\eqn\zy{
4z'' -  K^2  e^{8z}
 - e^{-\P + 4z - 4y-4w} ( {K'^2\ov 4 P^2} +
 e^{2 \P + 8 y+8w} P^2)
  - 12  H^2  e^{10y-4z }   =0\ ,
}
\eqn\ff{
(e^{-\P + 4z - 4y-4w} K')' - 2P^2 K e^{8z} =0 \ ,
}
with the first order  constraint
$$5  y'^2    - 2 z'^2  - 5 w'^2 - { 1 \ov 8} \P'^2
- { 1 \ov 4}  e^{-\P +  4z -4y - 4 w }{K'^2\ov 4 P^2}  $$
\eqn\coln{  -  \ { 3 } H^2  e^{10y -4z}
 -    e^{8y} ( 6 e^{-2w} - e^{-12 w} )
 +  { 1 \ov 4} e^{\P+  4z + 4y + 4 w } P^2 +
 { 1 \ov 8}  e^{8z} K^2   = 0  \ .
}
Lacking the exact solution of the above system, it was nonetheless
demonstrated in \bt\ the existence of a smooth interpolation
(in radial coordinate only)
 between (i)
 a non-singular
 short-distance region where the 10-d background  is
 approximately   $AdS_5\times T^{1,1}$ written in the
 coordinates where the $u={\rm const}$ slice
 is $S^4$, and
 (ii) a long-distance region  where the 10-d  background
 approaches the KT solution.
This was shown by  starting  with the short-distance
($u=\infty$ or $\r=0$) region,
i.e.    $AdS_5 \times T^{1,1}$
space (with the radius determined by the effective charge $K_*$)
and demonstrating   that  by doing perturbation theory in
the small parameter
${P^2\ov K_*} \ll 1$  one can match it onto  the KT  asymptotics
at large distances ($u\to 0$ or $\r \to \infty$).
The crucial point was  that $O({P^2\ov K_*})$
perturbations were regular at small distances.
One can literally repeat this analysis of \bt\ in our case to argue that
for  large enough Hubble parameter $H$ the naked singularity of the
KT geometry will be resolved. Here the short distance
region is a direct product of approximately
a de Sitter slicing of $AdS_5$  (as in section 2) and a $T^{1,1}$
coset.

The de Sitter deformation of the KT model above and of the LST 
in the previous section are similar 
in that as one turns off a Hubble parameter (or rather sufficiently 
lower it), one ends up with  
a singular geometry. 
A way to turn on small (vanishingly small)
de Sitter deformation  is to start with a gauge-gravity correspondence 
for a confining gauge theory like, say, the KS model \KS\ . 
It is straightforward to repeat above analysis for the deformed KS background 
and obtain a consistent system of equations. 
We do not present this system here due to its complexity and the fact
that we could not find  analytical solution. The added difficulty 
(compare to the extremal KS background) comes from the fact that 
it is inconsistent (on the level of equations of motion) to 
demand $H\ne 0$ along with a constant dilaton. A similar phenomenon 
has been observed in studies of the near extremal deformation of the KS 
background \btun\ : it was shown there that a black hole with 
a regular Schwarzschild horizon in the 
KS geometry necessarily has nonconstant  
dilaton.  This observation has a simple physical interpretation. 
In the extremal KS solution the string coupling $g_s$ was an exact modulus 
of the cascading gauge theory, dual to the sum of the individual 
gauge couplings 
\eqn\dils{
{1\over g_s}={4\pi\over g_1^2}+{4\pi\over g_2^2}={\ \rm const}\,.
}
As both the finite temperature and the Hubble parameter 
breaks supersymmetry, this modulus is expected to be lifted, thus 
developing a nontrivial radial dependence in the  dual supergravity.

%%%%%%%%%%%%%%%%%%%%%%%%
\newsec{Concluding remarks}
%%%%%%%%%%%%%%%%%%%%%%%%%%%%%%

In  this paper we  presented a simple framework how one can embed
an accelerating  universe in the supergravity. The idea is  to start
with a gauge-gravity duality of Maldacena,  and consider deformations
of this duality where  Minkowski background space-time of the gauge theory
is replaced with a de Sitter space-time.

We argued that to get nontrivial time-dependent solutions (i.e.
unrelated by coordinate reparametrization to a static solution)
the starting point for the deformation must be a gravitational
dual to a non-conformal  gauge theory. We discussed two 
examples of such deformations: the little string theory and the
KT model. In the both cases  conformal invariance is broken
by considering (adding) NS5 (D5) branes.  
We argued that the expansion of the background geometry on the gauge theory
side serves as an infrared cutoff in the dual supergravity.
In particular, for a sufficiently high expansion rate  this resolves
a naked singularity of the KT solution \KT\ .

There are several interesting future directions.
The vacuum state in an accelerating universe has a nonzero Gibbons-Hawking 
temperature  $T_{GH}=H/2\pi$, analogous to the Hawking temperature of a 
black hole.  The
KT deformation discussed here is very similar to the 
finite temperature deformation of the KT solution
considered in \refs{\fiv,\ghkt}. By comparing
a critical expansion rate for the $H\ne 0$ deformation
of the KT model with the critical temperature for its finite temperature
deformation, one should be able to relate the  Gibbons-Hawking temperature
of the expanding universe with the temperature of the gauge theory
in the standard near extremal deformation.

Another interesting question
is a dynamical stability of the deformed backgrounds 
discussed here. Since de Sitter deformation breaks supersymmetry
one has to worry about potential tachyons. The similarity of this 
deformation with the near-extremal one suggests that the KT model is likely 
to be stable, while there could be a tachyon in the LST deformation,
in analogy with \kuts\ . It would be nice to explicitly 
verify these conjectures. In the case of the KS deformation, at least 
for small values of the Hubble parameter, we expect to get 
a stable nonsupersymmetric background.  The argument is identical to 
the one given in \ass\ : the original supergravity 
background had a mass gap, and thus a small deformation 
should not produce a tachyon.

In this paper we {\it only} constructed  de Sitter backgrounds
in supergravity. It is  important to understand  the 
spectrum of density  fluctuations and the physics of D-brane probes in these 
geometries. 

Recently Giddings, Kachru and Polchinski \pgk\ studied embedding 
of the KS model in the type IIB string compactifications 
in the context of moduli stabilization and generation 
of large hierarchies of physical scales. It would be very interesting to 
explore de Sitter deformation of these models.

\bigskip
%\noindent
{\bf Acknowledgments}
%%%%%%%%%%%%%%%%%%%%%%%%%%%%%%%

\noindent
I would like  to thank Gary Horowitz, Amanda Peet, 
Joe Polchinski and Erich Poppitz for
useful discussions. I am especially  grateful to Arkady 
Tseytlin for valuable comments and reading the manuscript.  
This work is  supported in part
by the NSF under Grant No. PHY97-22022 and PHY99-07949.

%%%%%%%%%%%%%%%%%%%%%%%%%%%%%%%%%%%%%%%%%%%%%%%%%%%%%%%%
\vfill\eject
\listrefs
\end